\documentstyle[epsf,epsfig]{elsart}
\input{psfig}
\begin{document}
\begin{frontmatter}
\hoffset -.68in
\voffset -1.0in
\textwidth 6.5in
\textheight 9.0in
\title{
The Atmospheric Neutrino Flavor Ratio from a 3.9 Fiducial Kiloton-Year
Exposure of Soudan 2}
\author[3]{W.W.M. Allison},
\author[4]{G. J. Alner},
\author[1]{D. S. Ayres},
\author[3]{G. Barr\thanksref{a}},
\author[6]{W. L. Barrett},
\author[2]{C. Bode},
\author[2]{P. M. Border},
\author[3]{C. B. Brooks},
\author[3]{J. H. Cobb},
\author[4]{R. J. Cotton},
\author[2]{H. Courant}, 
\author[2]{D. M. Demuth}, 
\author[1]{T. H. Fields\thanksref{b}},
\author[3]{H. R. Gallagher},
\author[4]{C. Garcia-Garcia\thanksref{c}}, 
\author[1]{M. C. Goodman}, 
\author[2]{R. Gran}, 
\author[1]{T. Joffe--Minor},
\author[5]{T. Kafka}, 
\author[2]{S. M. S. Kasahara},
\author[1]{W. Leeson}, 
\author[4]{P. J. Litchfield}, 
\author[2]{N. P. Longley\thanksref{d}},
\author[5]{W. A. Mann}, 
\author[2]{M. L. Marshak}, 
\author[5]{R. H. Milburn}, 
\author[2]{W. H. Miller}, 
\author[2]{L. Mualem},
\author[5]{A. Napier}, 
\author[5]{W. P. Oliver}, 
\author[4]{G. F. Pearce},  
\author[2]{E. A. Peterson},
\author[4]{D. A. Petyt},
\author[1]{L. E. Price}, 
\author[2]{K. Ruddick}, 
\author[5]{M. Sanchez},  
\author[5]{J. Schneps},  
\author[2]{M. H. Schub\thanksref{e}},  
\author[1]{R. Seidlein\thanksref{f}}, 
\author[3]{A. Stassinakis}, 
\author[1]{J. L. Thron}, 
\author[2]{V. Vassiliev}, 
\author[2]{G. Villaume}, 
\author[2]{S. Wakely}, 
\author[5]{D. Wall}, 
\author[3]{N. West}, 
\author[3]{U. M. Wielgosz}
%                     
%$^{}$ \\
%\vskip 0.50cm
%
%
\maketitle
%
%\address{
\address[1]{Argonne National Laboratory, Argonne, IL
60439, USA }
\address[2]  {University of Minnesota, Minneapolis, MN
55455, USA }
\address[3]{ Department of Physics, University of Oxford,
Oxford OX1 3RH, UK }
\address[4]  {Rutherford Appleton Laboratory, Chilton, Didcot,
Oxfordshire OX11 0QX, UK }
\address[5]{ Tufts University, Medford, MA 02155, USA }
\address[6]  { Western Washington University, Bellingham, WA 98225, USA }

\thanks[a]  {Now at CERN, Geneva, Switzerland}
\thanks[b]  {Now at Fermilab, Batavia, IL 60510, USA}
\thanks[c]  {Now at IFIC, E-46100 Burjassot, Valencia, Spain  }
\thanks[d]  {Now at Physics Department, Colorado College, Colorado Springs,
CO  80903, USA}
\thanks[e]  {Now at Cypress Semiconductor, Minneapolis, MN, USA}
\thanks[f]  {Now at Lucent Technologies, Naperville, IL 60566, USA}
\begin{abstract}   

We report a measurement of the atmospheric neutrino flavor 
ratio, R, using a sample of quasi-elastic neutrino interactions 
occurring in an iron medium.  The flavor
ratio  (tracks/showers) 
of atmospheric neutrinos in a 3.9 fiducial kiloton-year exposure of Soudan 2 is 
$0.64\pm0.11\mathrm{(stat.)} \pm 0.06 \mathrm{(syst.)}$
of that expected.      
Important aspects of our main analysis have been checked
by carrying out two independent, alternative analyses; one is based 
upon automated scanning, the other uses a multivariate 
approach for background subtraction.  
Similar results are found by all three approaches.  

\end{abstract}
\end{frontmatter}
% typeset front matter (including abstract)

\section{Introduction}                 

The flavor ratio in sub-GeV atmospheric neutrino interactions 
as measured in underground detectors has sensitivity to a 
breakdown of the Standard Model in the neutrino sector.  
The flavor double ratio, defined as 
\begin{equation}
\mathrm{R_t} = \frac{[(\nu_\mu + \overline{\nu}_\mu)/(\nu_e + \overline{\nu}_e)]_{Data}}
         {[(\nu_\mu + \overline{\nu}_\mu)/(\nu_e + \overline{\nu}_e)]_{MC}},
\end{equation}                        
factors out the dependence on the absolute flux and in principle provides
a measurement with small systematic errors.  In practice the pure $\nu_\mu$ 
and $\nu_e$ rates cannot be measured directly and the experiments measure
\begin{equation}
\mathrm{R} = \frac{[\mathrm {(tracks)}/\mathrm{ (showers)}]_{Data}}
         {[\mathrm {(tracks)}/\mathrm{ (showers)}]_{MC}}
\end{equation}
for the iron calorimeters or 
\begin{equation}
\mathrm{R} = \frac{[(\mu~\mathrm{ ring)}/(e~\mathrm{ ring)}]_{Data}}
         {[(\mu~\mathrm{ ring)}/(e~\mathrm{ ring)}]_{MC}}
\end{equation}
for the water Cherenkov detectors.
The measured values of R depend on the acceptance and misidentification
in each individual experiment and are thus 
not expected to be equal to each other
or to $\mathrm{R_t}$.  However a measurement different from 1.0 in any experiment
is evidence of an anomaly.  

Six experiments have reported results on R~
\cite{Kam,IMB,SuperK,NUSEX,Frejus,flavor_pub}.
These measurements suggest a value of R significantly lower than 
unity.  The highest statistics on this measurement come from the water
Cherenkov experiments, Kamiokande, IMB, and 
SuperKamiokande.  
Three iron calorimeter experiments, NUSEX, Frejus, and 
Soudan 2, have reported results.  Our previous result [6],
$\mathrm{R}=0.72 \pm 0.19$(stat.)$^{+0.05}_{-0.07}$(syst.) was based
on an exposure of 1.52 kton-years.   The confirmation of
the low atmospheric flavor ratio with good statistical 
significance in a calorimeter would provide additional evidence
that there is no significant unexpected 
source of systematic error in
water detectors.  In this paper we report a value of 
$\mathrm{R} =0.64\pm0.11\mathrm{(stat.)} \pm 0.06 \mathrm{(syst.)}$ 
obtained in a 3.9 fiducial kiloton-year exposure  
of Soudan 2. 

There are three stages involved in our analysis.
First a 
sample of contained events is identified.  These are then classified
for neutrino flavor.   Finally a background subtraction is made and a value
of R calculated.  Each of these stages, 
particularly the flavor classification,
could introduce bias into the flavor ratio measurement. 
We have therefore checked the procedure by 
using different analyses.  These analyses give consistent results and 
confirm the validity of our principal result.  

Sections 3-5 describe our principal result, the flavor ratio as 
measured in a sample of quasi-elastic interactions from a 3.9 fiducial 
kiloton-year 
exposure.   The main analysis relies 
on physicist scanning for the verification of containment and for the 
flavor determination.  
By performing an analysis in which computer 
programs largely replace the
scanning (the Automated Analysis), we have verified
that the  
main procedure does not introduce biases due to subjectivity in the scanning.  
We have also checked our method of background subtraction and R calculation
by an additional analysis in which an alternative method for 
background estimation is used (the Multivariate Analysis). 
These analyses are described in Section 6.  

\section{The Soudan 2 Detector}                        
The Soudan 2 detector is a 963 ton fine-grained gas tracking
calorimeter located in the Soudan Underground Mine State Park in  Soudan, 
Minnesota.  
  The detector currently operates
with 90\% live time and has been taking data since 1989.  
It consists of 224 iron modules, each 1 meter x 1 meter x 2.7 
meters in size, and occupies a volume 
8 meters wide x 5.5 meters high x 16.1 meters long.  
Each module has a mass of 4.3 tons.  
Ionization deposited in the plastic tubes of a module drifts in an electric field
to the faces of the module where it is detected by vertical
anode wires and horizontal cathode strips.  The third coordinate
of the charge deposition is determined from the drift time in the 
module.  
The calorimeter modules operate in proportional mode; the measured
pulse height is proportional to the ionization deposited in the 
tube.  Pulse height measurements are used for particle identification.  
More details of the module construction and performance 
can be found in References \cite{DJC_1,DJC_2}.

The detector is surrounded by a 1700 $\mathrm{m}^2$ active shield mounted on
the cavern walls.  The shield is designed to identify particles entering
or exiting the cavern.   It tags events associated with cosmic ray muons 
passing close to the detector.   It has a measured efficiency of 95\% for 
cosmic muons crossing a shield element.  The complete shield covers 
about 97\% of the total solid angle. 
Reference \cite{shield} contains more information about the shield.

The most recent 2.4 kton-yrs of data were obtained with a number of improvements
to the detector modules and shield.  Additional layers added to the shield 
provided a third layer of shield coverage for much of the floor and ceiling
sections.  

\section{Data Analysis} 

The data described in this paper come from a 3.9 fiducial 
kiloton-year exposure
taken between April 1989 and January 1998.  This corresponds to a 
total exposure of 4.8 kiloton-years. 
During this time 
some 100 million triggers were recorded.   

\subsection{Contained Event Selection}
In the initial stage of our data analysis a sample of contained
events is selected.  A contained event is one in which 
all tracks and the main body of any showers are located within 
the fiducial volume, 
defined by a 20 cm depth cut on all sides of the detector.  
All events are processed by software filters. 
Events that satisfy the filter criteria are then scanned by physicists
to finalize the containment selection.
Data events are interspersed with Monte Carlo events so that 
on an event by event basis the scanner does not know if he or she is scanning
Monte Carlo or data.  
The scanning is performed in two stages with three independent scans
carried out at each stage. 
The contained event selection is fully described in 
References \cite{flavor_pub,HRG_thesis}.
                
The Monte Carlo sample used in this analysis is 5.45 times the size 
of the expected neutrino sample.  
The Monte Carlo simulates neutrino interactions in the detector;
background processes are not simulated.  
The Monte Carlo detector simulation reproduces the actual 
performance of the Soudan 2 detector to a high degree of 
accuracy.  The real detector geometry is simulated, as are local 
variations in the detector performance, particularly pulse height
and drifting.   Background noise in the detector is included
by overlaying Monte Carlo events onto randomly initiated triggers
generated throughout the exposure of this data set.  
For the first 2.2 kton-yrs exposure the MC and data events were combined
prior to the scanning while for the latter 1.7 kton-yrs exposure 
Monte Carlo events were inserted into 
the data stream during data acquisition at the Soudan mine.  
Data and MC events
are analyzed identically at each stage of the data reduction.  

\subsection{Flavor Classification}                  
During scanning, events
are classified into one of three categories:  single track, single shower, and
multiprong.  The single track category is further subdivided into mu-like
tracks and protons as described in the following paragraph.  
The track and shower categories
include primarily $\nu_\mu$ and $\nu_e$ 
quasi-elastic scattering respectively; they are largely 
equivalent to the `single ring'
category in the water Cherenkov experiments.  
In addition to the lepton, events in these categories may contain
recoil nucleons at the vertex and/or small showers
from muon decay at track endpoints. 
Events with two or more particles (other then recoil nucleons)
emerging
from the primary vertex, or single track events which are charged pions
having visible scatters, are classified as multiprong. 

Proton tracks can be identified because they are 
straight and highly ionizing.  
All tracks are fitted to a straight line  
trajectory and the track residual and average pulse
height are calculated.  Tracks with low fit residuals and high 
average pulse height are classified as protons.  
There is some overlap between protons and short, low energy muons where most
of the observed track has $\beta << 1$.  The separation 
algorithm is tuned to 
minimize the incorrect tagging of muons as protons.  Muon tracks are
incorrectly classified 4\% of the time and 80\% of protons are correctly 
identified.  
\begin{figure}[t]  
\vspace*{0.5in}
\epsfig{file=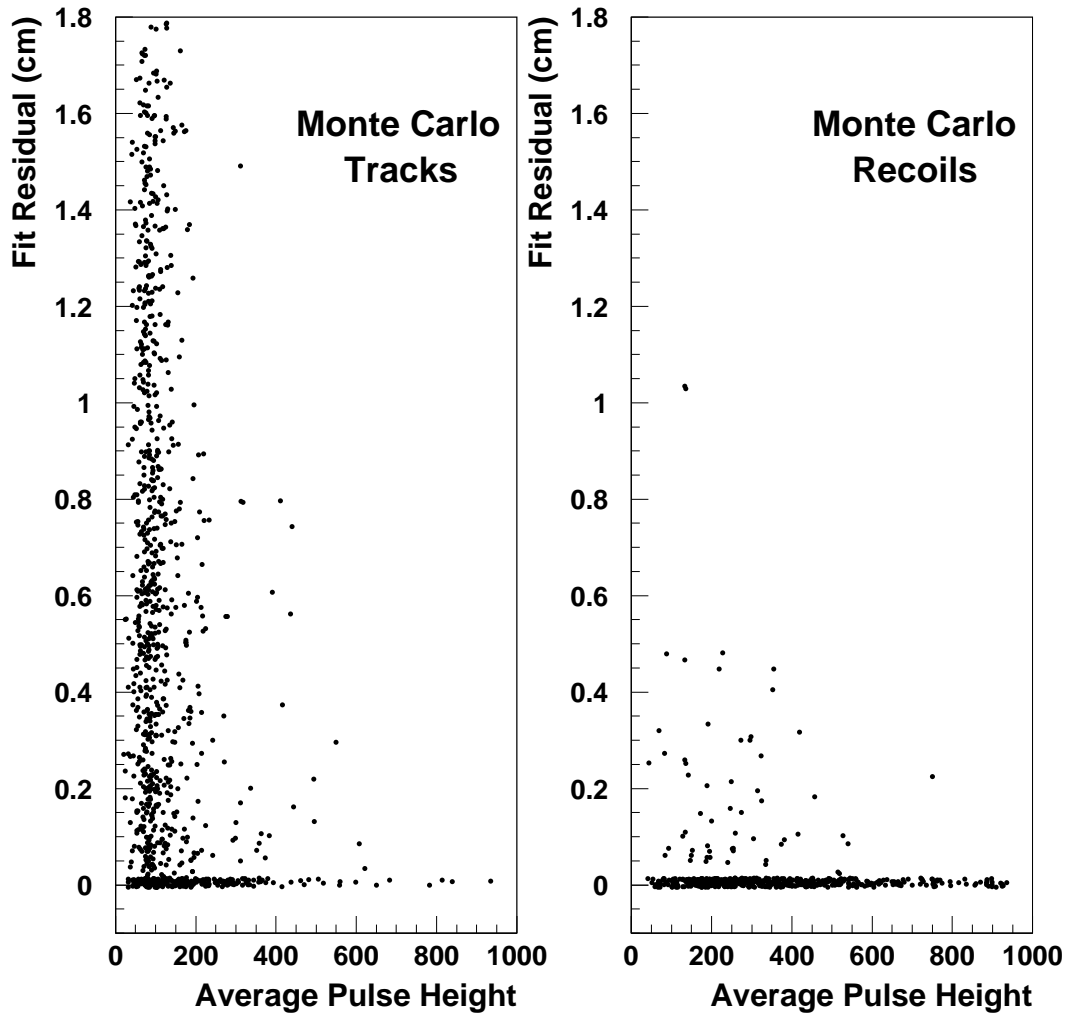,width=6in}
\caption{\label{proton} Proton Identification.  Fit residual vs. average pulse 
height for Monte Carlo events.  The left plot shows the results from 
2-d fits to the track in the event (usually a muon).  The right plot shows 
the results from the fit to the recoil (usually a proton) in track + 
recoil events.  The band at low fit residual in each plot is due to protons
and the band at low average pulse height to muons.}
\end{figure}        
Figure \ref{proton} shows the track residual 
versus average pulse height for MC tracks (mostly muons) and recoils
(mostly protons).

\section{The Flavor Ratio}         

\subsection{Shield Classifications}
Contained events
are a mixture of 
neutrino interactions and background processes.  
Neutral particles which are produced by the interaction of cosmic ray muons
in the rock surrounding the detector cavern are the principal source
of background.  
These
particles (neutrons and photons) can produce contained events if
they travel into the fiducial volume of the detector before interacting. 
Such events are usually accompanied by large numbers
of charged particles which strike the active shield located at the cavern
walls.  The presence of shield activity accompanying a contained event
therefore provides a tag for background events. 

The shield information allows us to identify two separate 
event samples in our data.  
An event with zero shield hits is referred to as `gold'; such 
an event is a neutrino candidate.  
Events with two or more shield hits are
referred to as `rock' events; they comprise a shield-tagged background
sample.     Table 
\ref{raw_numbers} gives the number of `gold', `rock', and Monte Carlo
events in each of the scanned categories.   Data events with one shield
hit are a mixture of neutrino events with a random shield hit, stopping
muons that pass the containment tests, and multiple shield hit events 
where shield hits are missing due to  shield inefficiency.  
Consequently the single shield hit events include both neutrino signal
and backgrounds, and we have excluded them from our analysis.  
The loss of neutrino events due to random shield hits is simulated 
by rejecting Monte Carlo events with shield hits in the randomly 
triggered background event.  

\subsection{Background Corrections}
\begin{figure}[t]  
\vspace*{0.5in}
\epsfig{file=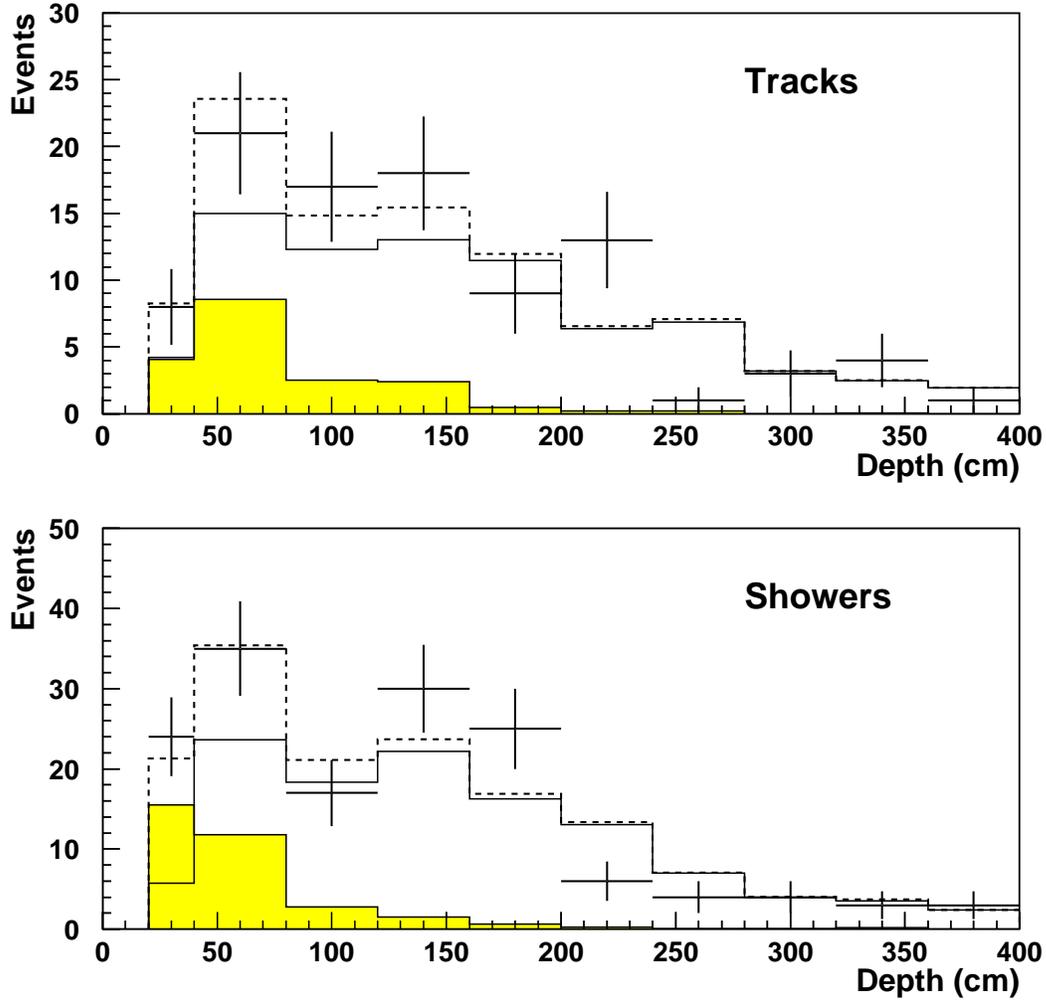,width=6in}
\caption{\label{depth_plots} Depth distributions.  Gold data are crosses.  
The rock distributions (shaded histograms) are normalized to the amount 
of background present in the gold sample as determined by the depth fit.
The MC distributions (open histograms) are normalized to the number of 
neutrino events present in the gold data as determined by the depth fit.
The dashed histogram shows the best fit to the data.  }
\end{figure}

It is possible that some muon interactions in the rock produce contained
events {\em unaccompanied} by shield hits,  due either 
to shield inefficiency or because the interaction 
did not produce any charged particles which entered the shield.  
The number of such interactions is  determined by examining the 
distributions of event depths in the detector, 
where the event depth is defined as the minimum distance
between the event vertex and the detector exterior, excluding the 
detector floor.  
These are shown in  
Figure \ref{depth_plots}.   

\begin{table}[ht]
\begin{center}
\begin{tabular}{lrrrr}       
\hline                   
Event & Track & Shower & Multi- & Proton \\ 
Type &  &  & prong &  \\     
\hline
Gold & 95 & 151 & 125 & 49 \\
Rock & 278 & 472 & 232 & 277 \\
MC & 749 & 729 & 711 & 82 \\
\hline
\end{tabular}
\caption
{Raw numbers of gold, rock (shield-tagged background) and
Monte Carlo 
events in each of the 4 categories.  }
\label{raw_numbers}
\end{center}           
\end{table}

We fit the depth distributions to determine the amount 
of background present in the gold sample.  
An extended maximum likelihood (EML) fit, that correctly handles bins
with small numbers of events,  is performed which describes the 
data distributions as
a sum of background and neutrino distributions.  
The shapes of the neutrino and background depth distributions are 
obtained from the MC simulation and the rock samples respectively. 
From the rock sample, we have determined that the track/shower ratio 
for background is $0.59\pm0.04$.
We have previously shown that the 
track/shower ratio of the background does not vary as a function of 
shield hit multiplicity \cite{flavor_pub}.  
We therefore expect background present in the gold (zero shield hit) 
sample to occur 
in this same track/shower ratio and we include
this expectation
as a constraint in the fit. 
Since the `flavor ratio' in the background is very similar to that measured
for neutrino events, background subtraction does not produce
a large change in our measured ratio. 

Early data had some contamination of the shower sample from electrical 
breakdown inside the modules.  A cut requiring at least 9 hits on 
all showers was used to remove this contamination.  A minimum of 6 hits
were required on all tracks.  

\section{The Flavor Ratio}                        
The results of the depth fits are that 
$76.9 \pm 10.8$ of the gold tracks and
$116.3 \pm 12.8$ of the gold showers are due to neutrino interactions.
We use these numbers to calculate the 
background corrected atmospheric neutrino flavor 
ratio.  Table \ref{results} shows the flavor ratios with and without 
(`raw') the background subtraction.  

\begin{table}[tb]
\begin{center}
\begin{tabular}{lr}
\hline
Number of Gold Tracks & 95 \\
Number of Gold Showers & 151 \\
Number of MC Tracks & 749 (137.4) \\
Number of MC Showers & 729 (133.8) \\ 
Corrected Number of $\nu$ Tracks & $76.9\pm10.8$ \\
Corrected Number of $\nu$ Showers & $116.3\pm12.8$ \\
& \\
Raw Value of R & $0.61 \pm 0.09$ \\
& \\
Corrected Value of R & $0.64 \pm 0.11$ \\
\hline
\end{tabular}
\caption
{Data used in the calculation of the corrected flavor ratio.   The 
Monte Carlo numbers in parentheses are normalized to the detector
exposure. The error on the flavor ratio is statistical only.}
\label{results}
\end{center}           
\end{table}                                 

The systematic error 
due to the background subtraction has two components.  

\begin{enumerate}
\item 
Many of the single tracks in the background sample are protons, 
which come from neutrons entering the fiducial volume of the 
detector and elastically scattering. 
Hence the proton classification, which removes single protons from 
the track sample and places them in a separate category, serves as 
a background correction even before the depth fits are performed.  
An alternative approach to the one we have taken is to leave the 
single protons in the track sample and determine the amount of background
solely from the depth fits. 
The resulting value for the flavor ratio differs from our main value 
by $\delta \mathrm{R} = +0.023$; the full difference is taken as a component of 
the systematic
error.    

\item 
Our method assumes that any background 
present in the gold sample behaves 
identically to the shield-tagged background of the rock sample.  We have
investigated how the results change if this assumption is not valid.  
For instance, if the zero-shield hit shower background has 
a different depth distribution than the rock shower sample then the depth 
fit (which assumes the rock distribution for the background) will 
incorrectly estimate the amount of background present.  We have considered
a number of such effects and have determined the resulting uncertainty 
on R to be $\delta \mathrm{R} = \mbox{}^{+0.041}_{-0.030}$.  

\end{enumerate}
The total error on R due to the background subtraction 
can be obtained by adding these two contributions in quadrature:  
$\delta \mathrm{R} = \mbox{}^{+0.047}_{-0.038}$.
Systematic errors due to the uncertainty in the expected flavor ratio, 
Monte Carlo, and scanning procedure are calculated to be 
$\delta \mathrm{R} = \pm0.040$ \cite{flavor_pub}.  Here we have taken the 
systematic uncertainty in the expected flavor ratio to be 
$\delta \mathrm{R/R} =3\%$.  Adding this in quadrature to 
the error from the background subtraction results in a total systematic
error of 
$\delta \mathrm{R} = \mbox{}^{+0.062}_{-0.056}$.
Our primary measurement of the flavor ratio is therefore
R = $0.64\pm0.11 \mathrm{(stat.)} \pm 0.06 \mathrm{(syst.)}$.   

\section{Alternative Analysis Methods} 
We have used two other methods to determine the flavor content
of the atmospheric neutrino flux.  

\subsection{Multivariate Discriminant Analysis} 
This method uses the same event sample as the main analysis but uses
additional event variables to provide 
discrimination between background and neutrino interactions.  
There are several quantifiable differences between background events and 
neutrino events.  In comparison with neutrino interactions, background 
events are (on average) closer to the detector exterior,
of lower energy, traveling preferentially downward, and 
are more likely to be traveling into the detector (rather than outward).
The corresponding variables are the event depth, energy, zenith angle,
and `inwardness' (defined as the cosine of the angle 
between the event direction and the inward-pointing normal vector to 
the nearest face of the detector). These event
variables 
are
combined, using the method of multivariate discriminant 
analysis~\cite{MVA_ref,JS_pdk}, to 
form a single variable.  

\begin{figure}[t]  
\epsfig{file=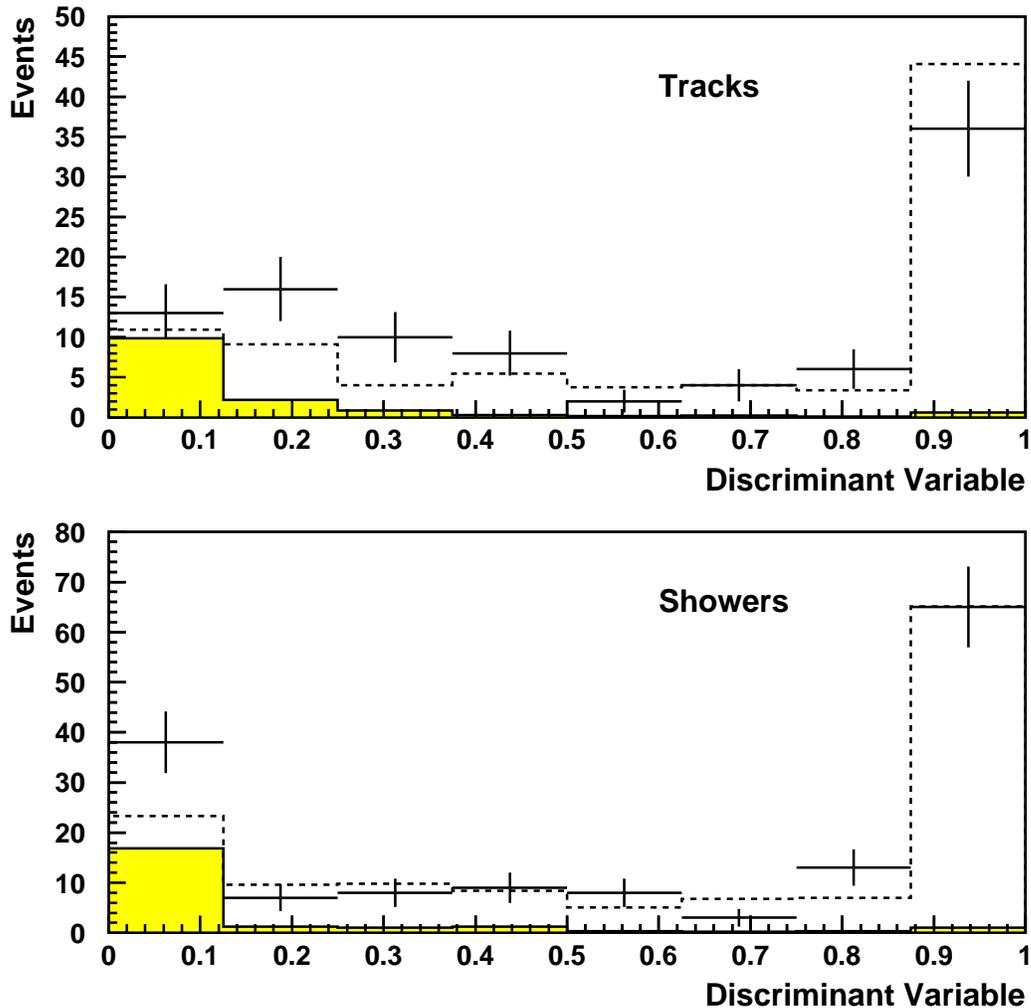,width=6in}
\caption{\label{MVA} 
Distributions of the multivariate discriminant variable 
combining event depth, energy, zenith angle, and inwardness.  
Track distributions are at top, showers at bottom.  Gold data are crosses.
The rock distributions (shaded histograms) are normalized to the amount 
of background present in the gold sample as determined by the fit
to the multivariate distributions.
The MC distributions (open histograms) 
are normalized to the number of 
neutrino events present in the gold data as determined by the fit to 
the multivariate distributions. }
\end{figure}

Figure \ref{MVA} shows the distributions of the discriminant 
variable for MC neutrino, rock, and gold events.
The improved discrimination between the expected neutrino and 
rock distributions
is clearly visible in the first and last bins.  
A fit of the 
discriminant variable distributions for the gold events to a sum of the
distributions for rock and Monte Carlo events then gives the amount of
background for track and shower events separately and thus a value of R.
The track/shower ratio in the background sample is again used to provide
a constraint on the zero shield hit background determined from the fit, 
as in the main analysis. 
The result of the fit, 
using the four variables described above is $\mathrm{R}=0.61\pm0.10$ (stat. only). 
                                                     
Unlike the event depth, the distributions of
other variables in this analysis
would be influenced by possible new physics.  This analysis 
therefore should only be regarded as a test of the null hypothesis, i.e. no 
new physics, for which the value expected is R=1.0.  

\subsection{Automated CEV Selection and Flavor Determination} 
\begin{figure}[t]  
\epsfig{file=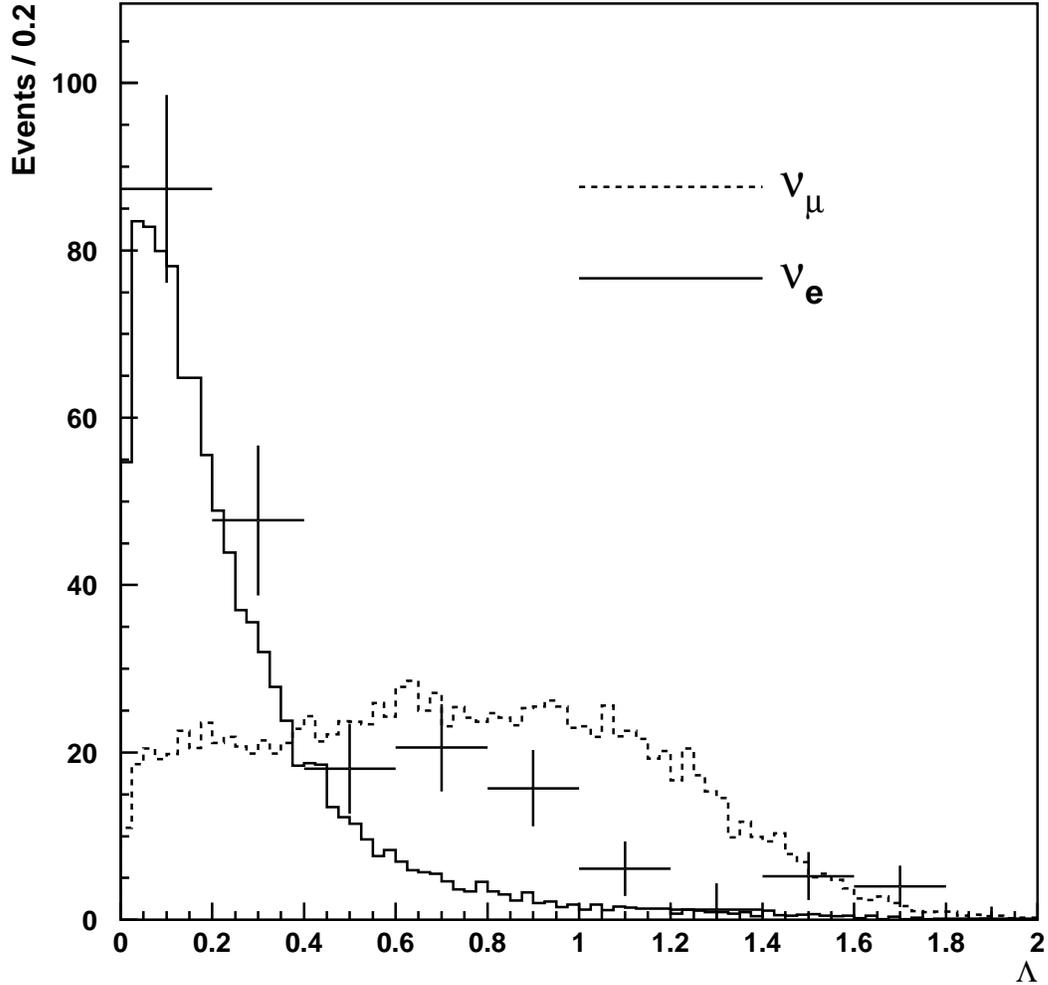,width=6in}
\caption{\label{lambda} Distributions of $\Lambda$ for $\nu_\mu$ (dashed 
histogram) and $\nu_e$ (solid histogram) Monte Carlo events which 
survive event selection cuts.  All events produced by the given neutrino 
flavor (charged and neutral current) are included.  Background corrected
GOLD data are shown as crosses.
The deficit of $\nu_\mu$-like events relative
to $\nu_e$-like events in the data is clearly evident.  } 
\end{figure}

A third analysis that uses software for both event selection and 
flavor determination has been developed and 
is described in detail elsewhere~\cite{AS}.  
It has been applied to a subsample of the data corresponding to a 
2.7 fiducial kiloton-year exposure.  
Since this method has been applied to a smaller data set and the contained
event selection is done differently, the event sample used in this analysis is
not identical to that used in the main analysis.  
 The important virtues of this method are that it 
almost entirely eliminates the role of subjective 
decisions and permits the use of much larger Monte Carlo samples.  The 
shield is used to separate data events into neutrino candidates (GOLD
events) and 
a background sample (ROCK events), using the same principles as in the main 
analysis.  The `GOLD' and `ROCK' samples here are not the same as the gold and rock 
samples 
described previously since the contained event samples are different and 
shield information is used differently.  
Events are required to have more than 10 hits. 
The distinction between track and shower events is made on the 
basis of $\Lambda$, a variable derived from a spherical harmonic
analysis of inter-hit correlations in each event.  Figure \ref{lambda} 
shows distributions of $\Lambda$ 
for the data and all true $\nu_\mu$ and $\nu_e$ interactions as given 
by the Monte Carlo.  
All events are included and no attempt is made to 
distinguish neutral current interactions, inelastics and quasi-elastics.
From the Monte Carlo we expect neutral current interactions to account for 
12\% of the events in the final data sample.
         
Events are separated
into `$\nu_\mu$-like' and `$\nu_e$-like' sub-samples by application of an
energy dependent cut 
on $\Lambda$.  Of $\nu_\mu$ interactions (neutral and charged 
current), 78.3\% are correctly tagged as $\nu_\mu$-like while 80.7\% of 
$\nu_e$ interactions are correctly tagged as $\nu_e$-like.   

A depth fit is performed to the GOLD data in terms of a combination 
of ROCK and Monte Carlo distributions~\cite{AS}.  
The results of this procedure are summarized in Table \ref{AA_results}.
The ratio of ratios is $\mathrm{R}=0.62\pm0.14$(stat.)$\pm0.05$(syst.).
This figure is calculated using all events: quasi-elastic, inelastic and
neutral current.  

\begin{table}[tb]
\begin{center}
\begin{tabular}{|l|c|c|}
\hline
& $\nu_\mu$-like & $\nu_e$-like \\
\hline
Monte Carlo & 17353 &15915 \\
\hline
Rock & 328 & 410 \\
\hline
Contained Events & 109 & 171 \\
due to $\nu$-interactions & $74.9\pm13.5$ & $111.5\pm15.7$ \\
due to Rock & $34.1\pm12.1$ & $59.5\pm14.4$ \\ 
\hline
\end{tabular}
\caption
{Effects of the flavor cut on the Monte Carlo, rock, and data 
distributions in the Automated Analysis.}
\label{AA_results}
\end{center}           
\end{table}                                 

\section{Conclusion}                 

The flavor ratio of atmospheric neutrinos (data/MC) has been measured
from a 3.9 fiducial kiloton-year exposure of the Soudan 2 detector to be
$0.64\pm0.11\mathrm{(stat.)}\pm0.06\mathrm{(syst.)}$.   
This result is obtained after applying a background correction to a 
sample of 246 quasi-elastic neutrino 
candidates.
The probability of a statistical fluctuation
to R=0.64 or below is less than $4\times 10^{-3}$.  
Two other independent analyses have been performed.
These check the contained event 
selection, flavor determination, and background correction procedures 
of our 
main analysis.  
Both alternative analyses confirm the validity of the main analysis.
The good agreement of these three rather different 
methods gives confidence that the effect is not an
artifact of a particular analysis.  
This measurement is in good agreement with the previously 
published result from 
this experiment as well as the results
from the water Cherenkov experiments.

\section*{Acknowledgements}
This work was undertaken with the support of the U.S. Department of Energy,
the State and University of Minnesota, and the U.K. Particle Physics and
Astronomy Research Council.  
 We would also like to thank:  the Minnesota
Department of Natural Resources for allowing us to use the facilities of 
the Soudan Underground Mine State Park; the staff of the Park, 
particularly Park Managers P. Wannarka and J. Essig, for their day to day 
support; and Messrs B. Anderson, J. Beaty, G. Benson, D. Carlson, J.
Eininger and J. Meier  
of the Soudan Mine Crew for their work in the 
installation and running of the experiment.

\end{document}